# GJ 3470 c: A Saturn-like Exoplanet Candidate in the Habitable Zone of GJ 3470

Phillip Scott[1], Bradley Walter[2], Quanzhi Ye[3], David Mitchell[4], Leo Heiland[5], Xing Gao[6], Alejandro Palado [7], Burkhonov Otabek[8], Jesús Delgado Casal [9], Colin Hill [10], Alberto García [11], Kevin B. Alton [12], Yenal Ogmen [13], Vikrant Kumar Agnihotri [14], Alberto Caballero[15]

11 July, 2020

[1] OKSky Observatory, OK, USA.
[2] Paul and Jane Meyer Observatory, TX, USA.
[3] California Institute of Technology, CA, USA.
[4] California Polytechnic State University, CA, USA.
[5] Heiland Observatory, USA.
[6] Xingming Observatory, Xinjiang, CN.
[7] Al Sadeem Observatory, UAE.
[8] Maidanak Observatory, UZ.
[9] Nuevos Horizontes Observatory, Spain.
[10] Iris Observatory, Australia.
[11] Rio Confio Observatory, Spain.
[12] Desert Bloom Observatory, AZ, USA.
[13] Green Island Observatory, Cyprus.
[14] Cepheid Observatory, India.
[15] Habitable Exoplanet Hunting Project, Spain.

**Abstract**

We report the discovery of a new exoplanet candidate orbiting the star GJ 3470. A total of three transits were detected by OKSky Observatory: the first one on December 23, 2019, the second one on February 27, 2020, and the third one on May 3, 2020. We estimate an average transit depth of 0.84 percent and duration of 1 hour and 2 minutes. Based on this parameter, we calculate a radius of 9.2 Earth radii, which would correspond to the size of a Saturn-like exoplanet. We also estimate an orbital period of 66 days that places the exoplanet inside the habitable zone, near the orbital distance at Earth's equivalent radiation. Another twelve potential transits that do not belong to GJ 3470 b are also reported. Despite our candidate for GJ 3470 c still has to be confirmed by the scientific community, the discovery represents a turning point in exoplanet research for being the first candidate discovered through an international project managed by amateur astronomers.

# 1    Introduction

The Habitable Exoplanet Hunting Project is the first international program coordinated by amateur astronomers to search for habitable exoplanets. The group includes more than 30 amateur and professional observatories located in more than 10 countries across the five continents (Caballero, 2020). When a new target is assigned, each observatory observes the same star during several months. By doing so, we increase the chances of discovering new exoplanets.

The main objective of the project is to search for nearby potentially habitable planets. For this reason, target hosts are selected from non-flaring G, K and M-type stars within 100 light years, having known transiting exoplanets outside the habitable zone or non-rocky exoplanets inside. As of June 2020, we have conducted observing campaigns on GJ 436, GJ 1214 and GJ 3470.

The campaign on GJ 3470 started the 20th of December, 2019, and finished the 15th of March, 2020. We were initially looking for potentially habitable exoplanets similar to Kepler 296-f, with a transit depth of 0.2 percent, duration of 3 hours, and orbital period of 6 weeks (Rowe, 2014). Our data showed three potential transits with an average transit depth of 0.84 percent, a duration of 1 hour 2 minutes, and orbital period of 66 days. This could be close to the orbital distance at Earth's Equivalent Radiation, but our observations did not rule out periods equal to the observed period divided by integers.

So far, only the exoplanet GJ 3470 b has been discovered orbiting our target star. GJ 3470 b is a Neptune-like exoplanet 13.9 times the mass of the Earth, 0.408 times the radius of Jupiter, and with an orbital period of 3.3 days (NASA, 2020).

If confirmed, GJ 3470 c would be the first exoplanet totally discovered by amateur astronomers. Some exoplanets such as KPS-1b have been discovered using data gathered by amateur astronomers. However, the discovery of KPS- 1b was made by the Kourovka Planet Search (KPS), a project run by Ural State Technical University.

# 2    Methodology

The method used to discover our exoplanet candidate is transit photometry, which detects changes in brightness of a star potentially caused by an exoplanet transiting between the star and our point of view. Our project has a series of guidelines that every observatory should implement. Some of them are, for example, to acquire at least two data points per minute, conduct observing sessions of at least four hours of duration, using the same comparison stars while analysing the data, and using AstroImageJ for the photometry. AstroImageJ is a popular software developed by University of Louisville for astronomical image analysis and precise photometry (University of Louisville, 2020). Figure 1 shows the comparison stars used in AstroImageJ, none of which are variable:

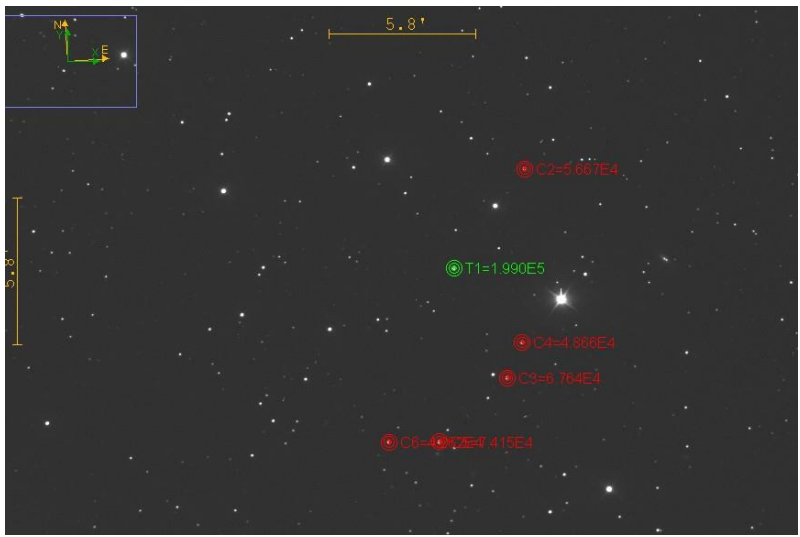

Figure 1 – Comparison stars of GJ 3470
Source: OKSky Observatory

We used multi-aperture photometry with a radius of object aperture of 5 pixels, an inner radius of background of 10 pixels, and an outer radius of background annulus of 15 pixels. Detrend for airmass is applied. The data below corresponds to our target star GJ 3470 [1]:

| Distance: | 30.7 pc |
|---|---|
| Spectral type: | M1.5 |
| Apparent magnitude: | 12.3 |
| Mass: | 0.51 Solar masses |
| Radius: | 0.48 Solar radii |
| Metallicity | 0.2 Fe/H |
| RA: | 07:59:06 |
| DEC: | +15:23:30 |

---
[1]Exoplanet.eu (2020)

All the images taken by OKSky Observatory were 60 second exposures with a luminance filter. This observatory is located at the OKSky Ranch, Kiowa, OK 74553 (coordinates -95:53:56, +34:41:24), altitude: 220 meters. The telescope used by the observer was a 12.5" f4.8 Newtonian with a GEM SiTech drive mount, and a SBIG ST-8XE CFW-8 camera.

## 3  Transit photometry

The three transits were detected and reported by the same observatory, OK- Sky, in OK, USA. The first transit occurred on December 23, 2019 (Tc: 2458840.656). The transit showed a depth of 0.93 percent, a duration of 51 minutes, and a radius of 0.93 Jupiter radii.

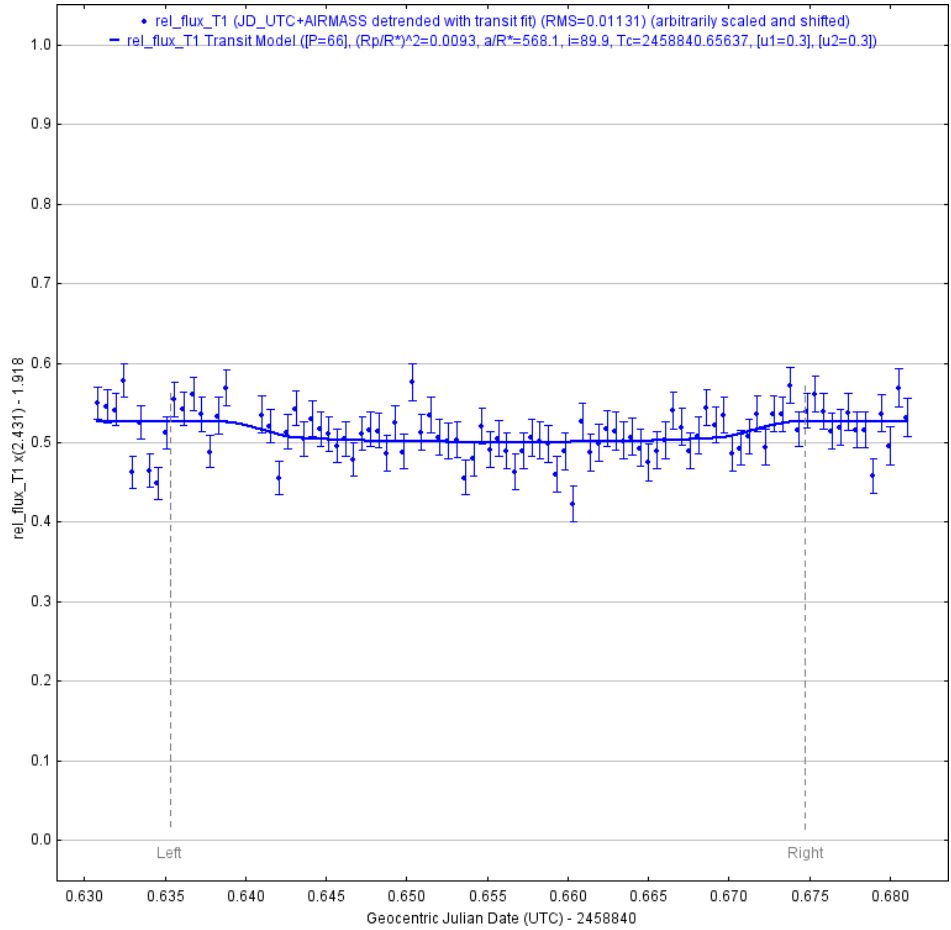

Figure 2 – First transit of candidate GJ 3470 c
Source: OKSky Observatory

The second transit was detected on February 27, 2020 (Tc: 2458906.778). It showed a depth of 0.55 percent, a duration of 1 hour and 5 minutes, and a radius of 0.73 Jupiter radii.

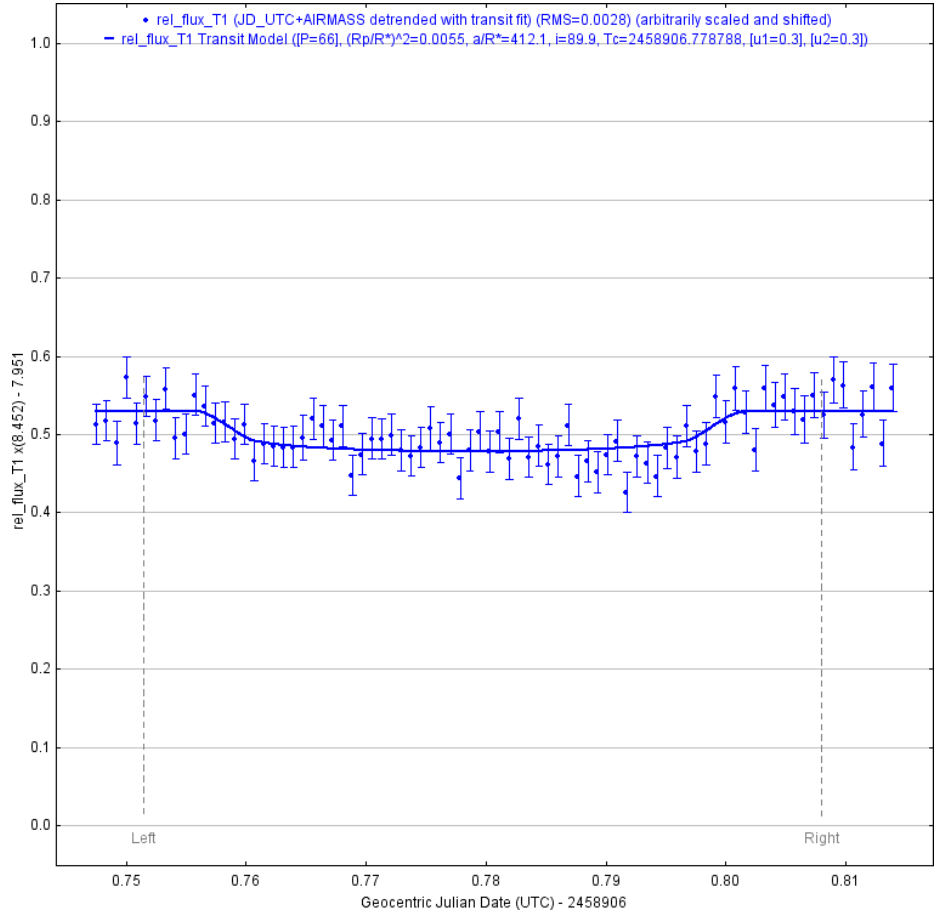

Figure 3 – Second transit of candidate GJ 3470 c
Source: OKSky Observatory

A third transit was detected on May 3 (Tc: 2458972.638). This last transit has an estimated depth of 0.79 percent and duration of 1 hour and 11 minutes, and a radius of 0.87 Jupiter radii.

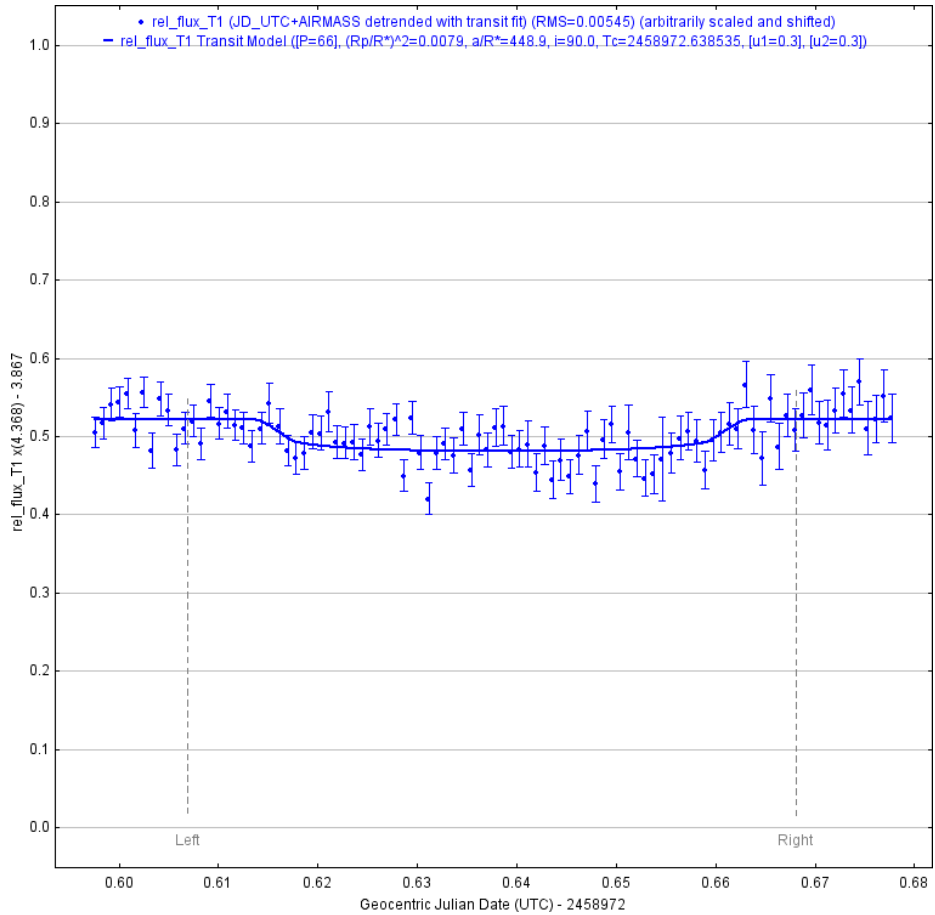

Figure 4 – Third transit of candidate GJ 3470 c
Source: OKSky Observatory

After reviewing the estimated transit times of GJ 3470 b in the NASA Exoplanet Archive, we have determined that our observed transits are incompatible with the known transiting exoplanet. We have also discarded them as a possible explanation because GJ 3470 b has a transit depth of 0.58 percent and a duration of 1.9 hours (Bonfils, 2012).

Moreover, on February 27, OKSky observatory was able to capture the transit of both GJ 3470 b and our candidate in the same light curve. Image below.

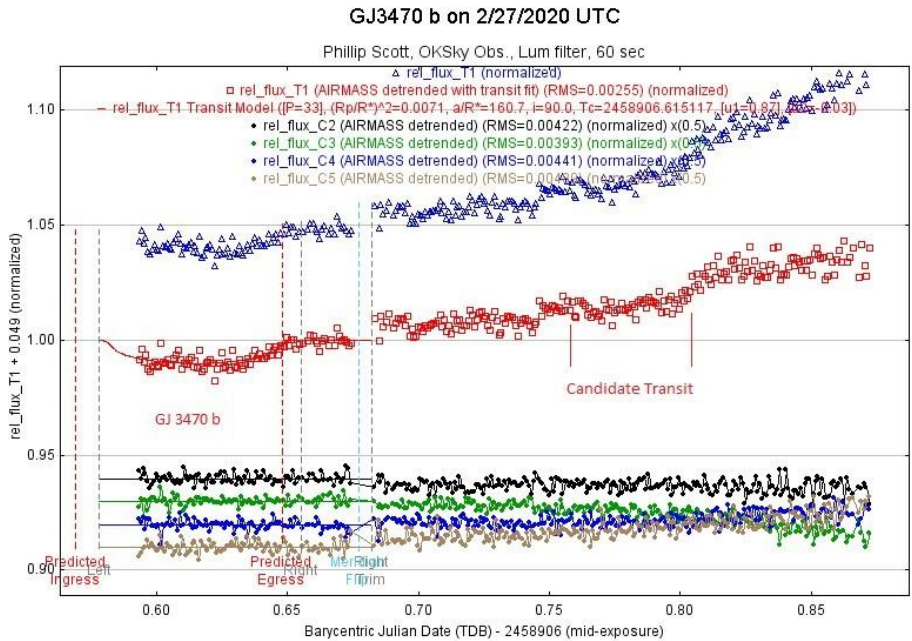

Figure 5 - Transits of GJ 3470 b (left) and our candidate GJ 3470 c (right)
Source: OKSky Observatory

## 4 Discussion

Our candidate has an observed orbital period of 66 days, but observations do not rule out periods related to the observed by integer divisors. Despite the goal of the project to monitor specific stars continuously, substantial gaps occurred due to technical issues, bad weather or other circumstances.

For example, the candidate could have an orbital period of 33 or 16.5 days. No observatory was able to observe the 8th of January, 10th of February, 14th of March or 16th of April, which are the days that the transit of an exoplanet with a shorter period could have taken place.

The main concern we had during the analysis of our data was to mistakenly identify a variation caused by star spots as a potential transit. Dragomir et. al (2012) has pointed out that unocculted spots in GJ 3470 only produce a difference in Rp/Rs of no more than 0.0004.

In the same line, Biddle et. al (2014) has indicated that a 1 percent peak-to-valley variability in GJ 3470 implies a time-dependent, spot-induced variability of $5 \times 10^{-5}$ over the rotation period.

This, together with the fact that the transit we detected in December is between 0.14 and 0.38 percent deeper and between 14 and 20 minutes shorter than the other two transits, lead us to believe that it could have been caused by a star spot. The following linear regression shows our three potential transits:

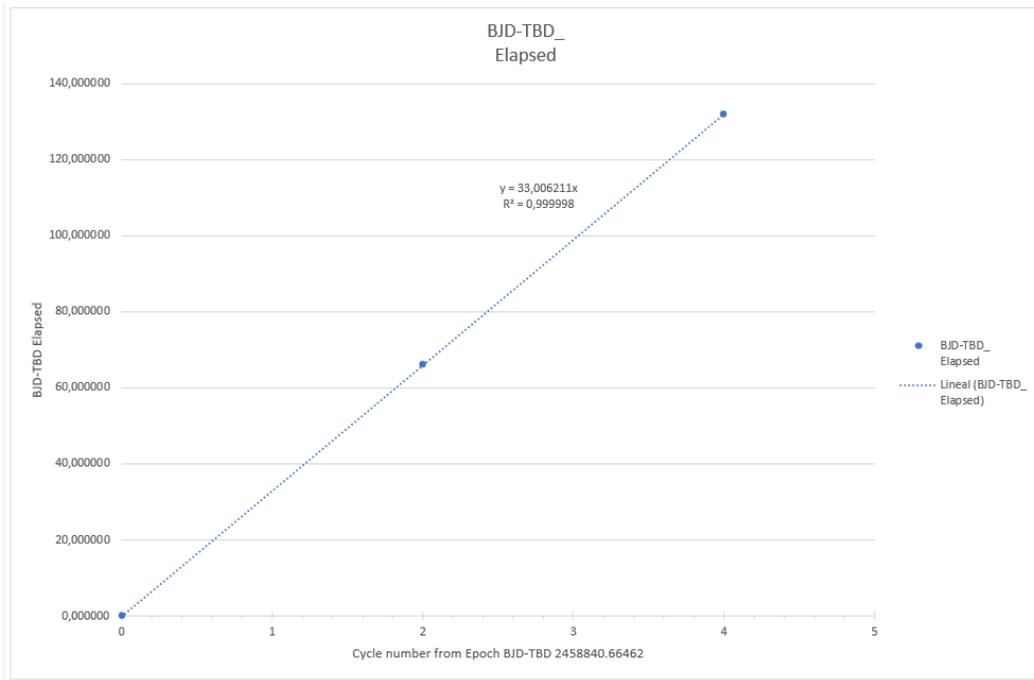

Figure 6 – Linear regression of GJ 3470 c candidate transits
Source: Paul and Jane Meyer Observatory

The following graph shows the three potential transits of our candidate:

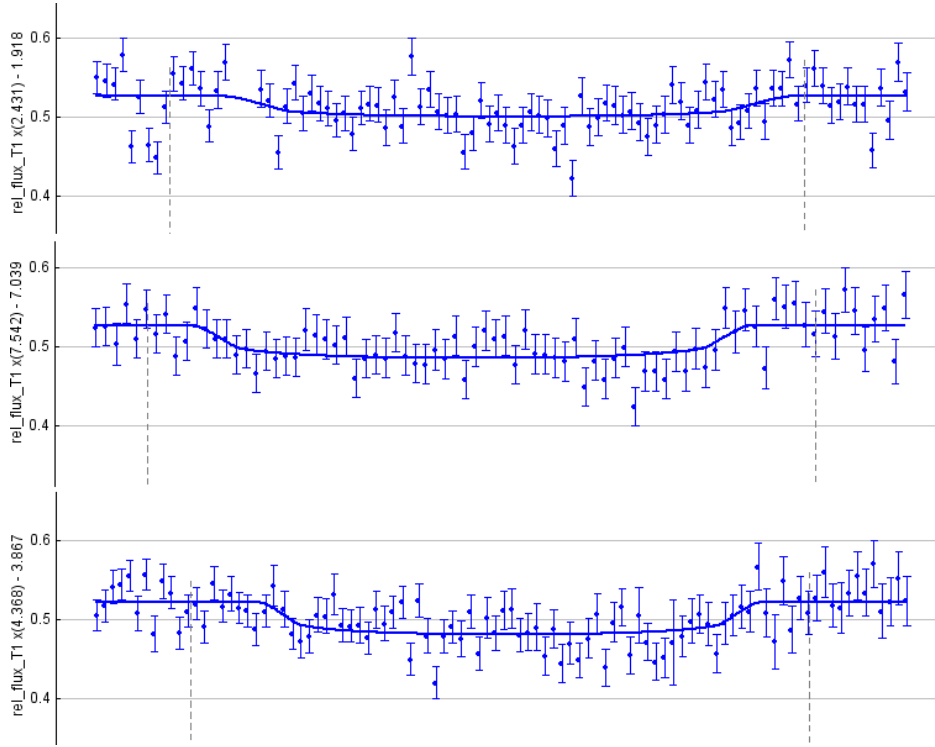

Figure 7 – Comparison of the three potential transits of GJ 3470 c
Source: OKSky Observatry

## 5 Characterization of candidate GJ 3470 c

The first parameter we wanted to know was the radius of our exoplanet candidate. The following equation gives a radius of 4.5 Earth radii.

$$D = (r/R)^2$$
$$r = R \cdot \sqrt{D}$$
$$r = 334,243 \times \sqrt{0.0075} = 28,946 \text{km}$$

*D = transit depth
*r = planet radius
*R = star radius

However, this equation would only be true if the star is an evenly illuminated disk. AstroImageJ takes into account limb darkening, estimating a radius for exoplanet of 0.84 Jupiter radii, which equals to 58,725 km, only 493 km more than the radius of Saturn.

The second parameter we looked at was the semi-major axis of our candidate. Considering a stellar mass of 0.51 solar masses and a period of 66 days, we obtained a semi-major axis of 0.25 AU, which falls closer to the orbital distance at Earth's Equivalent Radiation. The habitable zone of GJ 3470 extends from 0.139 (inner boundary) to 0.292 (outer boundary) (Exoplanet Kyoto, 2020).

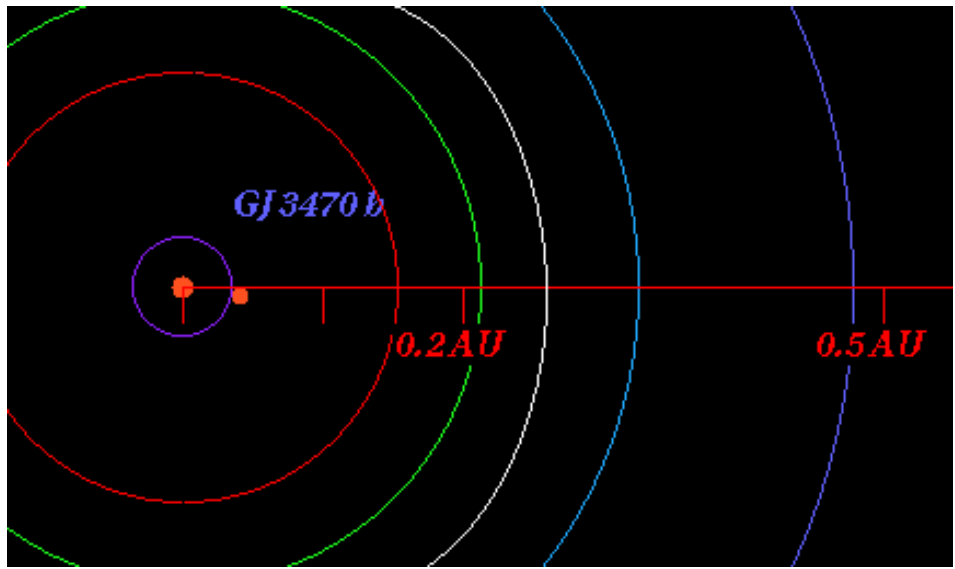

Figure 8 – Habitable zone of GJ 3470 (the white circle corresponds to the orbit of our candidate, and the green circle to the orbital distance at Earth's Equivalent Radiation)
Source: Exoplanets Kyoto

Our exoplanet candidate is unlikely to host any habitable exomoon. Habitable exomoons should orbit exoplanets with a mass of at least 3 times that of Jupiter, or 954 times the mass of the Earth (René, 2013). Based on the correlation between the radius and mass of exoplanets, we estimate a mass of around 100 Earth masses for our candidate.

## 6 Other candidates

We also detected twelve other transits that could belong to different exoplanets, but we were not able to estimate any orbital period. Iris Observatory detected a 1 percent transit on January 3, with a duration of 2 hours and 32 minutes and a radius of 0.99 Jupiter radii.

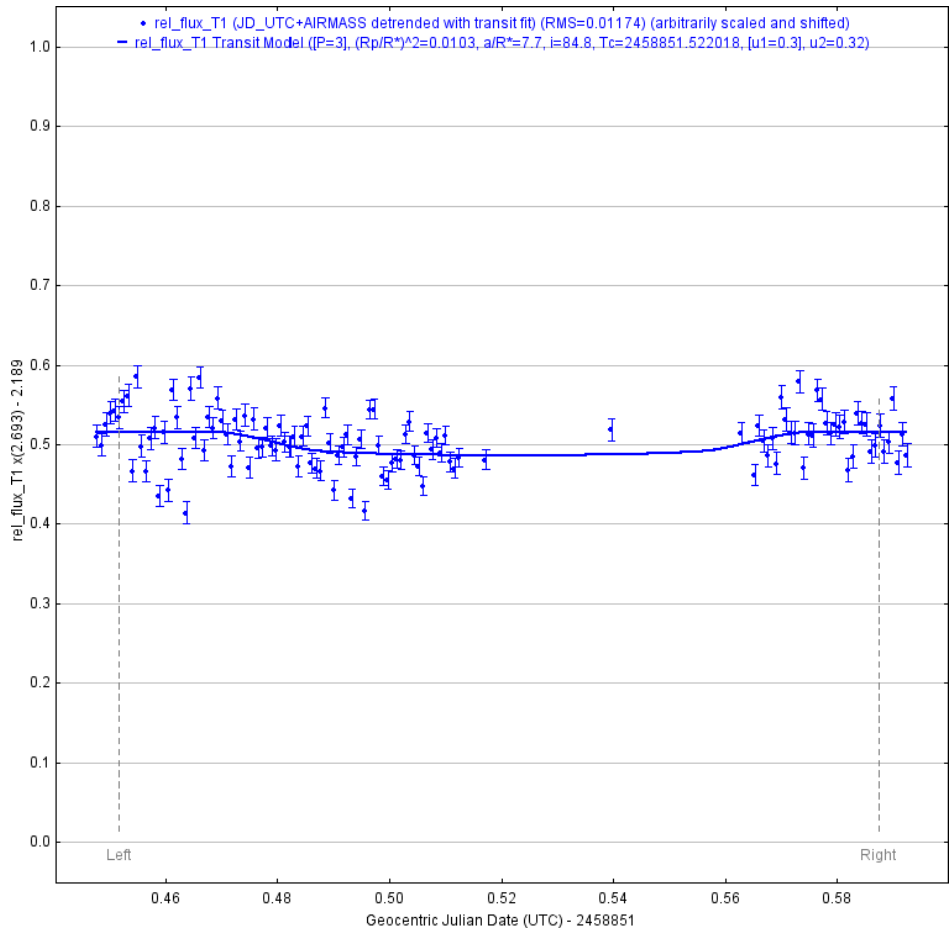

Figure 9 – Possible transit of new exoplanet
Source: Iris Observatory

On January 4, OKSky Observatory detected a transit with a depth of 1.2 percent, a duration of 3 hours and 10 minutes, and a radius of 1.1 Jupiter radii.

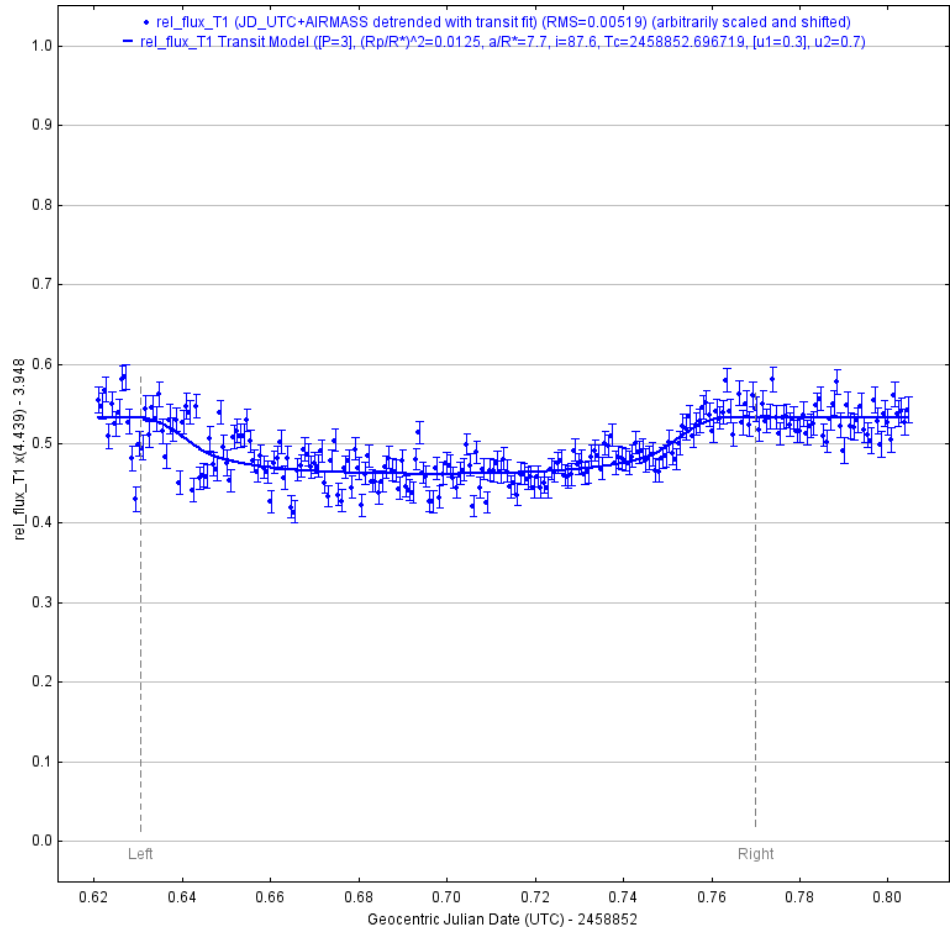

Figure 10 – Possible transit of new exoplanet
Source: OKSky Observatory

On January 9, Desert Bloom observatory detected a potential transit with a depth of 0.1 percent, a duration of 2 hours and 37 minutes, and an estimated radius of 0.15 Jupiter radii.

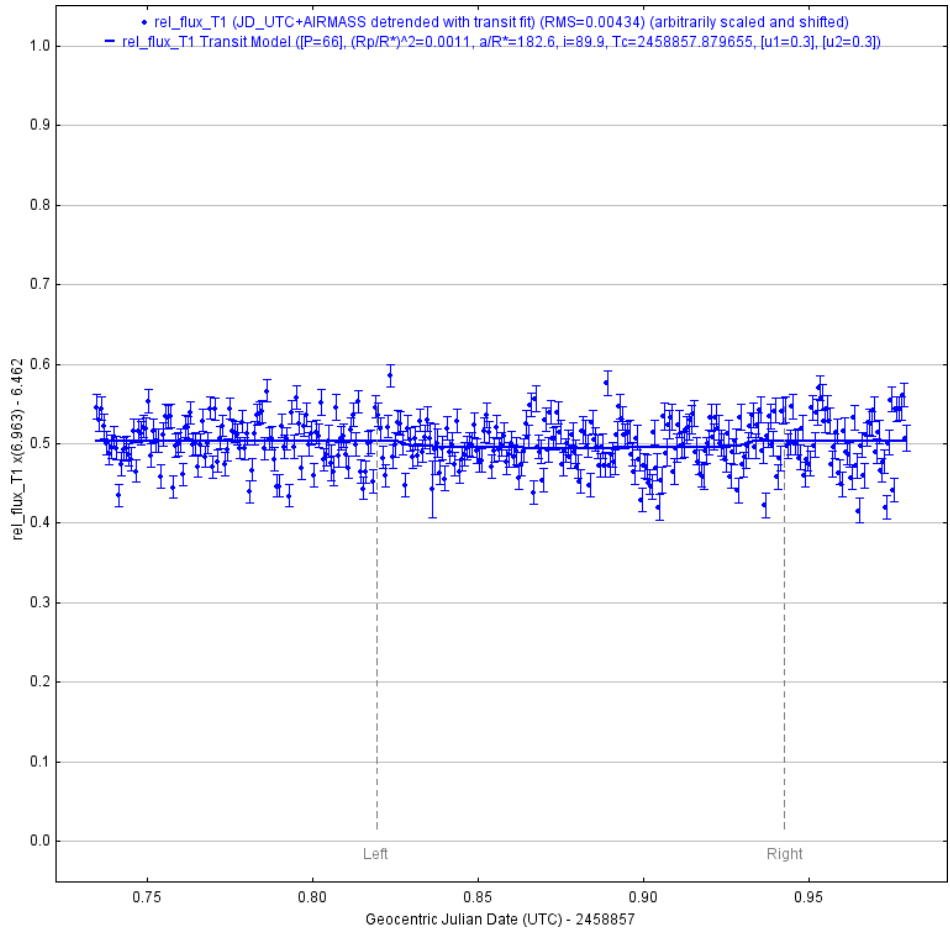

Figure 11 – Possible transit of new exoplanet
Source: Desert Bloom observatory

On January 17, Iris observatory detected a potential transit with a depth of 0.76 percent, a duration of 3 hours and 45 minutes, and an estimated radius of 0.86 Jupiter radii.

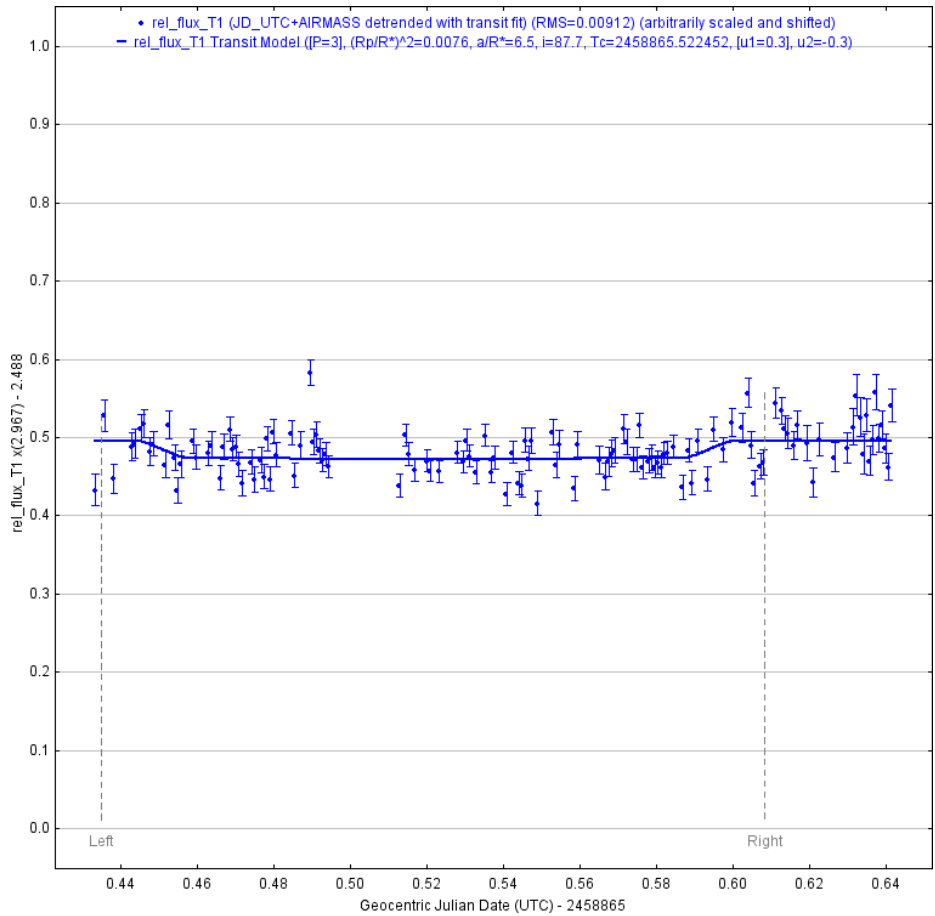

Figure 12 – Possible transit of new exoplanet
Source: Iris observatory

OKSky observatory also reported a third transit on January 19, with a depth of 0.47 percent, a duration of 2 hours and 22 minutes, and a radius of 0.68 Jupiter radii

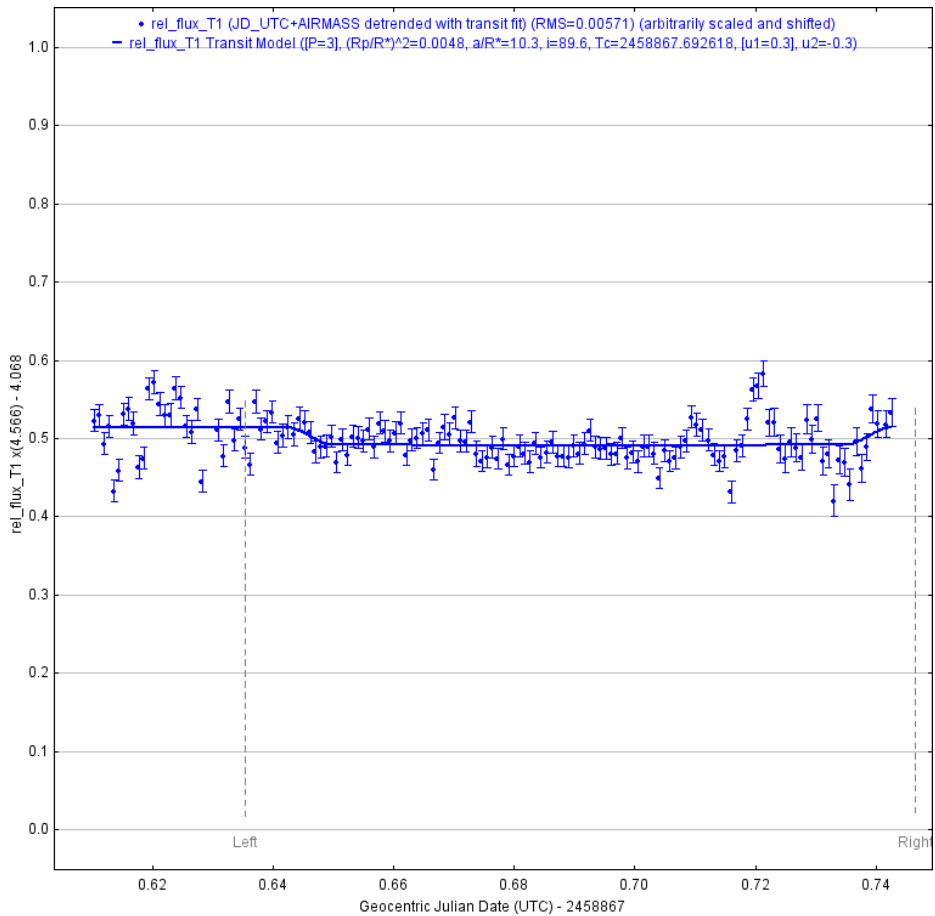

Figure 13 – Possible transit of new exoplanet
Source: OKSky Observatory

On January 25, OKSky Observatory captured two partial transits that could belong to two different new exoplanets: one with a depth of 2.6 percent and an estimated duration of 2 hours and 20 minutes, and the other one 5 percent, with an estimated duration of 4 hours.

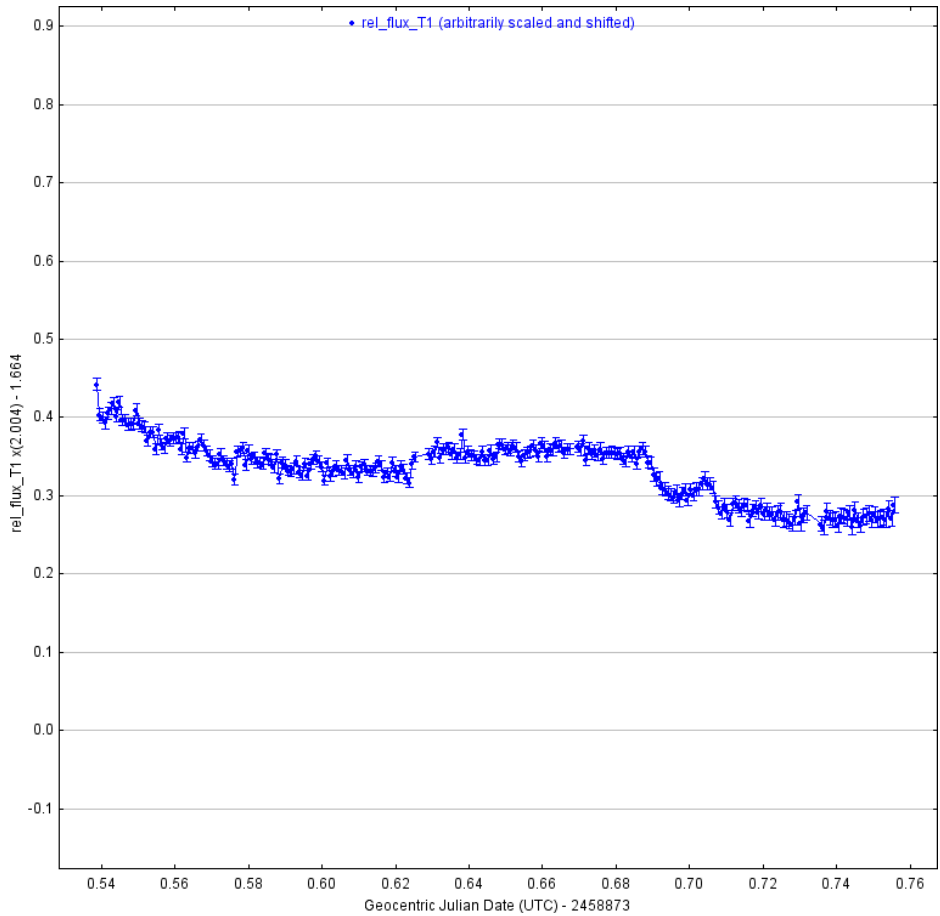

Figure 14 – Possible partial transits of two new exoplanets
Source: OKSky Observatory

On January 27, Paul and Jane Meyer Observatory detected a potential transit with a depth of 0.36 percent, a duration of 55 minutes, and an estimated radius of 0.28 Jupiter radii.

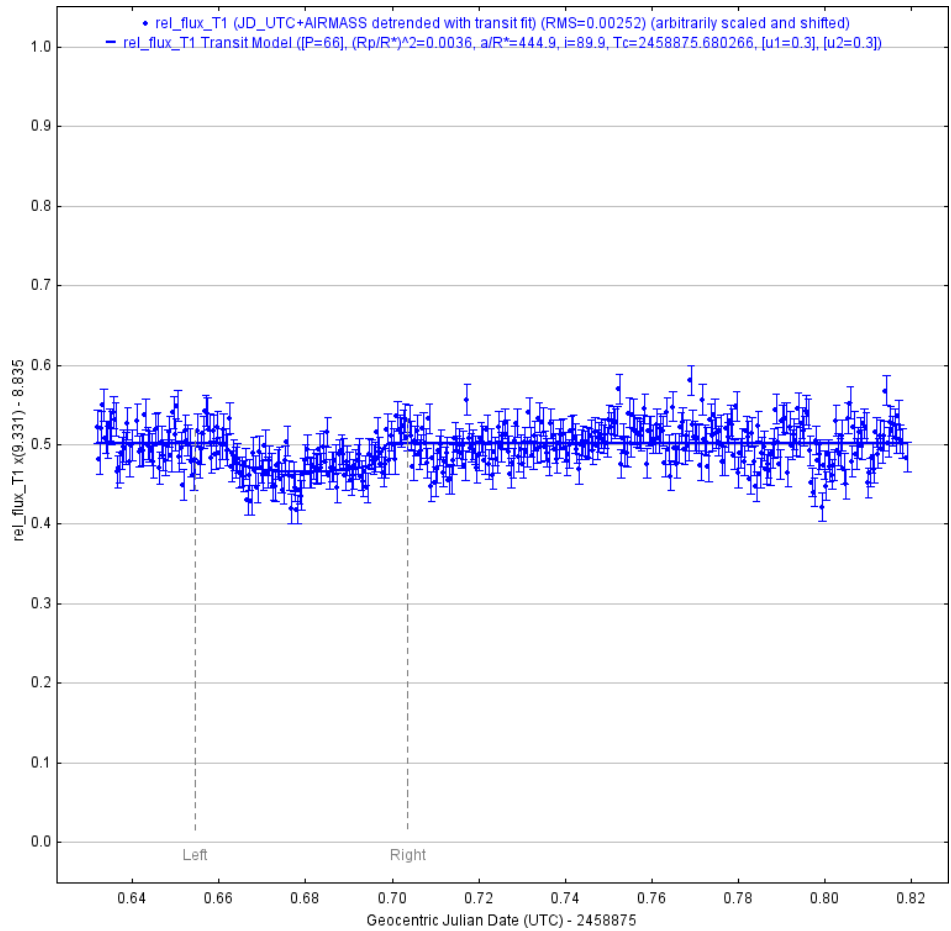

Figure 15 – Possible partial transits of new exoplanet
Source: Paul and Jane Meyer Observatory

On February 2, Paul and Jane Meyer Observatory detected a potential transit with a depth of 1.2 percent, a duration of 3 hours and 51 minutes, and an estimated radius of 1.1 Jupiter radii.

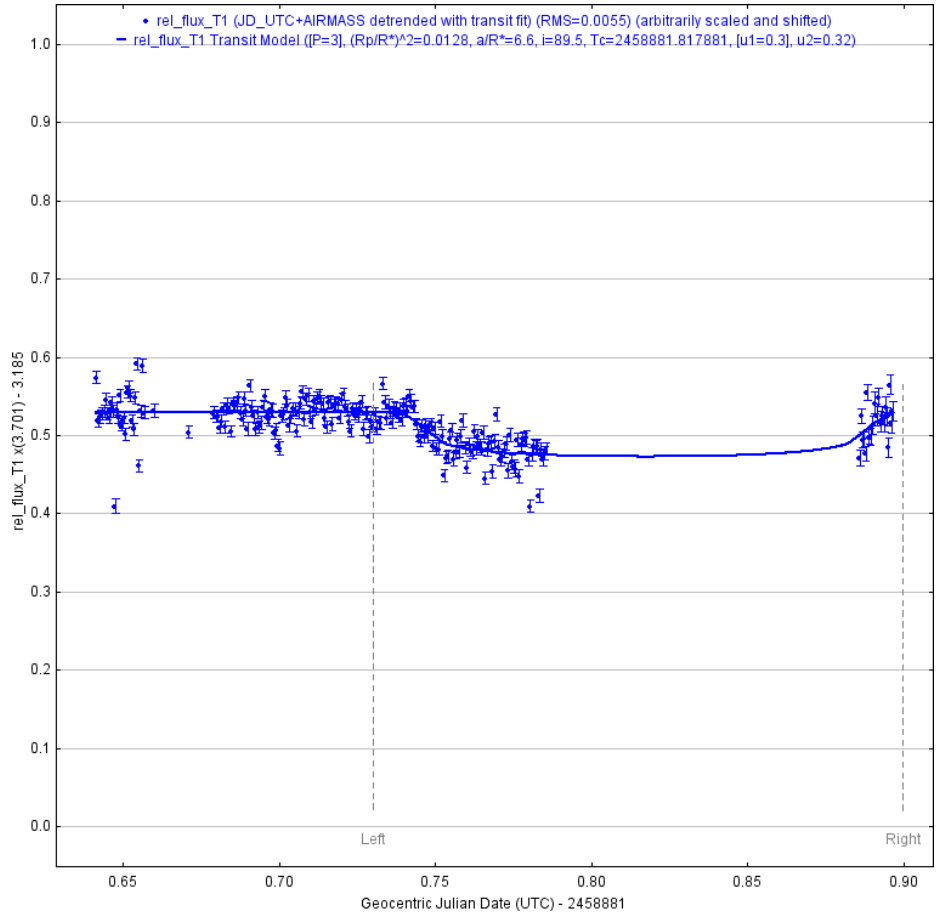

Figure 16 – Possible transit of new exoplanet
Source: Paul and Jane Meyer Observatory

This candidate resembles the 0.76 percent-transit detected by Iris Observatory on January 17, with a duration of 3 hours and 45 minutes, and an estimated radius of 0.86 Jupiter radii. These two detections are separated by 16 days.

Heiland Observatory in Arizona reported on February 18 a 0.46 percent transit with a duration of 1 hour and 57 minutes, and an estimated radius of 0.67 Jupiter radii.

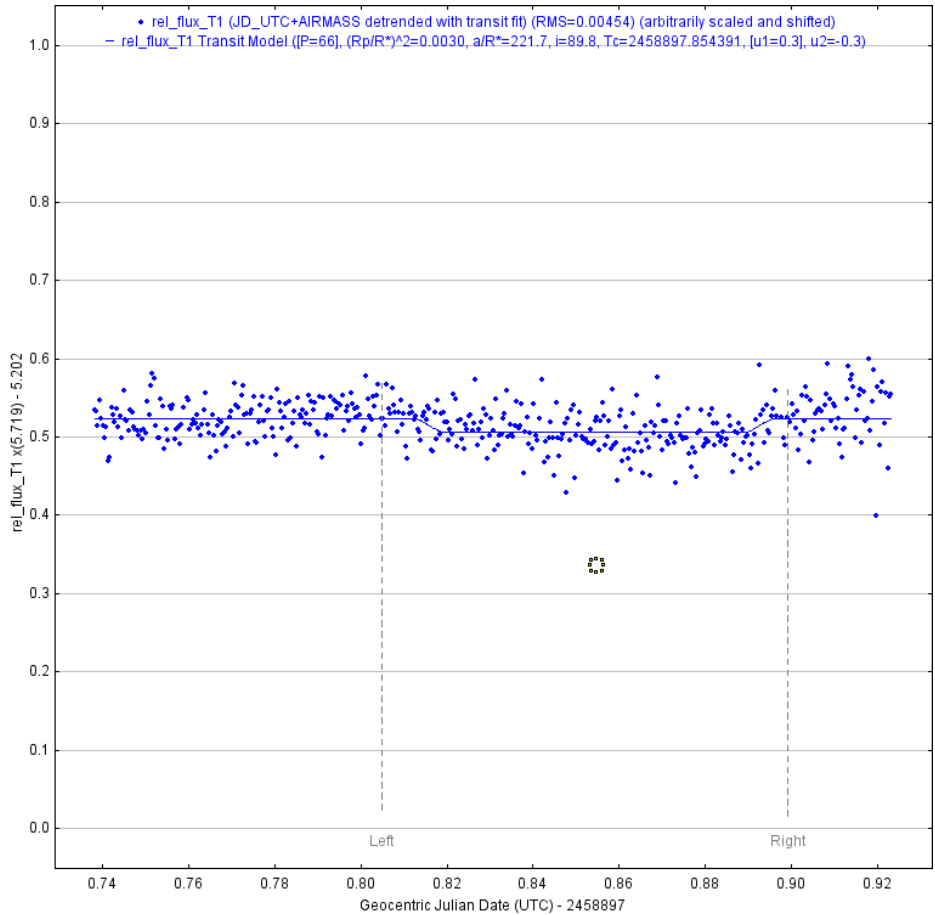

Figure 17 – Possible transit of new exoplanet
Source: Heiland Observatory

This candidate resembles to the 0.47-percent transit detected by OKSky observatory on January 19, with a duration of 2 hours and 22 minutes, and a radius of 0.68 Jupiter radii. These detections are separated by 30 days.

On April the 1st, OKSky Observatory reported a potential transit with a 1 percent depth and a duration of 1 hour and 2 minutes. The radius estimated by AstroImageJ was 0.48 Jupiter radii.

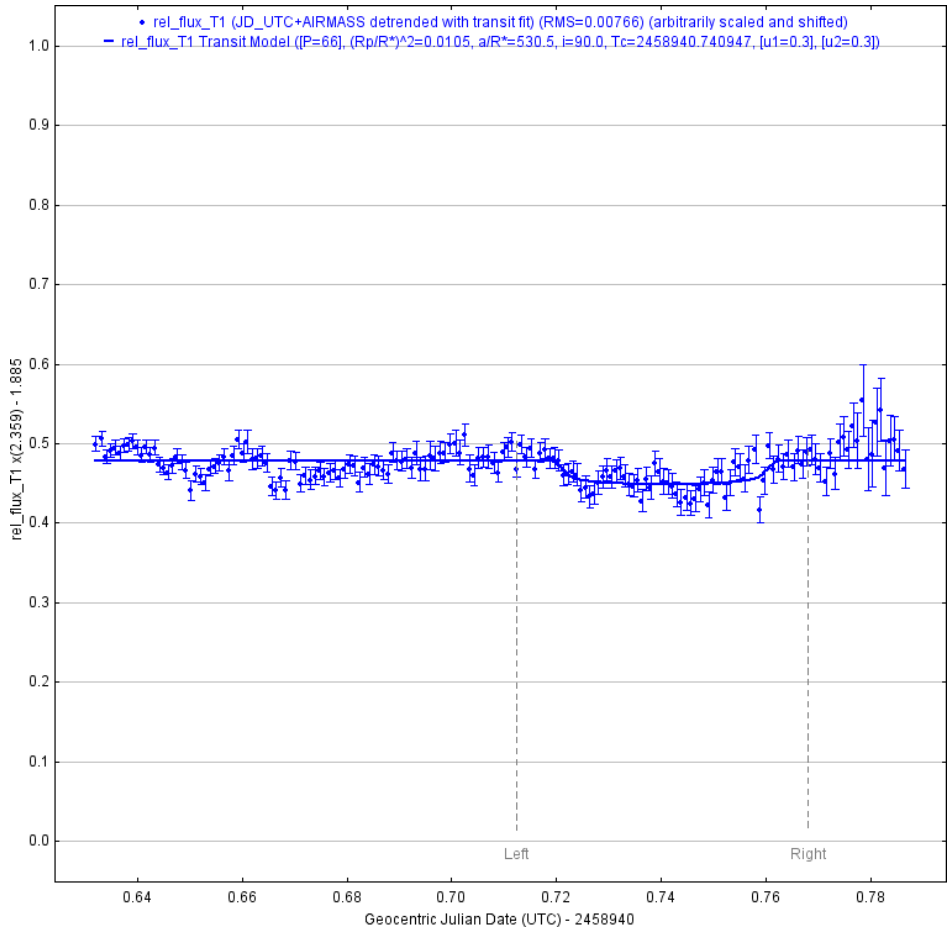

Figure 18 – Possible transit of new exoplanet
Source: OKSky Observatory

On April 26th, OKSky Observatory reported another potential transit with a depth of 1.5 percent, a duration of 2 hours and 11 minutes, and a radius of 0.57 Jupiter radii.

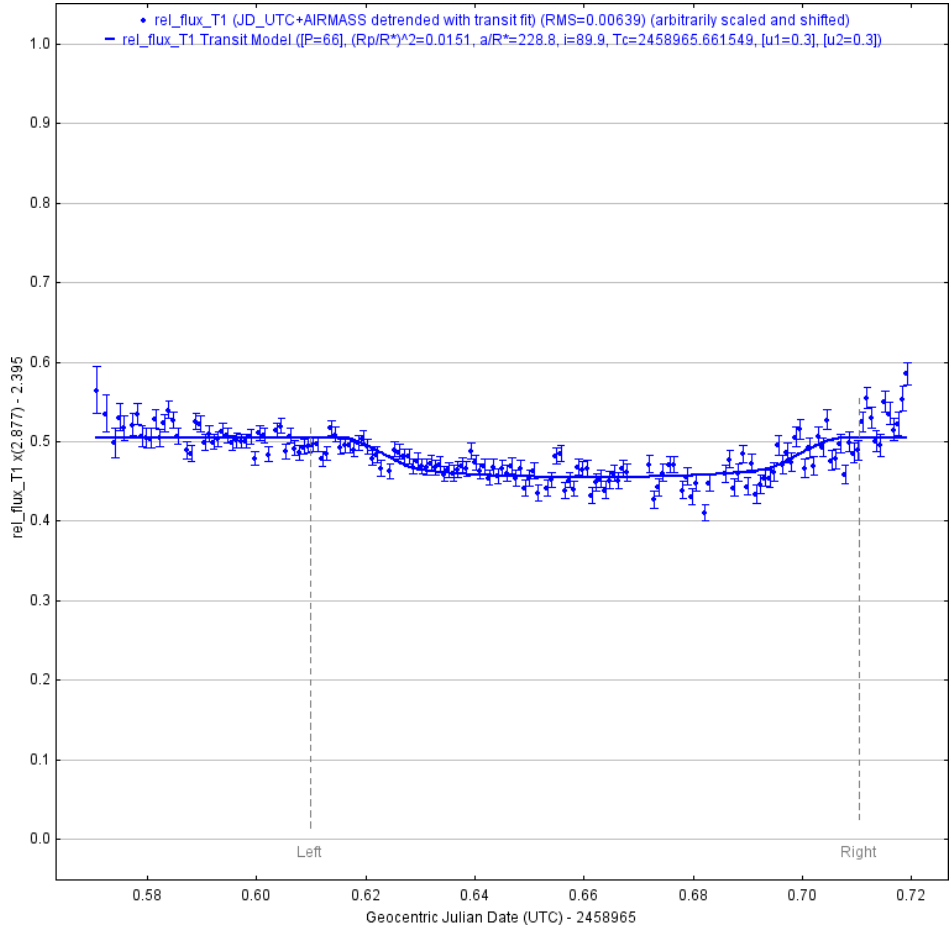

Figure 19 – Possible transit of new exoplanet
Source: OKSky Observatory

## 7   Conclusions

In this paper we report the detection of a new exoplanet candidate orbiting the red dwarf star GJ 3470. OKSky Observatory detected three transits: one on December 23, 2019, another one on February 27, 2020, and a final one on May 3, 2020. We estimate an average transit depth of 0.84 percent and a duration of 1 hour and 2 minutes. However, we suspect that the first transit we detected, in December, could have been caused by a star spot.

Based on our estimation of the transit depth, we calculate a radius of 9.2 Earth radii. This size would correspond to an exoplanet slightly larger than Saturn. We also calculated an orbital period of 66 days, which would place the planet inside the habitable zone of the GJ 3470 star system.

Twelve more potential transits were also reported. Out of these twelve detections, two of them seem to have repeated. The two transits could be caused by different gas giants, one of them with an orbital period of 16 days and the other one 30 days.

Our candidate for GJ 3470 c still has to be confirmed by other professional observatories, but the diversity of potential transits we found suggests a high likelihood of existing more exoplanets in the system apart from GJ 3470 b. The discovery represents a new approach to exoplanet research for being the first exoplanet candidate fully discovered by amateur astronomers.

## 8 Data

Our data can be downloaded through the following links:

December 23, 2020 transit candidate: https://drive.google.com/drive/folders/1ndfK93X0MwE8zQB5aUahfzv3cnkKwvTV?usp=sharing

February 27, 2020 transit candidate: https://drive.google.com/drive/folders/17Xi-gHJq38iTsFrtJbe5sLj7sxZ8LpUR?usp=sharing

May 3, 2020 transit candidate: https://drive.google.com/drive/folders/1cf6A-O5alLgnr53N2-pqTsJqZUbljght?usp=sharing

## 9 Acknowledgements

We thank the participation of several students who contributed to this project but were not included in the list of authors. They are Chaomin Chen, Margo Thornton, and Kailei Gallup, from California Polytechnic State University.